\documentclass[useAMS,usenatbib]{mn2e}
\usepackage{graphicx}

\title[A new substantive proton to electron mass ratio constraint on rolling scalar field cosmologies]
{A new substantive proton to electron mass ratio constraint on rolling scalar field cosmologies}
\author[Rodger I. Thompson]{Rodger I. Thompson$^{1}$\thanks{E-mail:
rit@email.arizona.edu (RIT)}\\
$^{1}$Steward Observatory, University of Arizona, Tucson, AZ 85721, USA\\}
\begin{document}

\date{Accepted xxxx. Received xxxx; in original form xxxx}

\pagerange{\pageref{firstpage}--\pageref{lastpage}} \pubyear{2013}

\maketitle

\label{firstpage}

\begin{abstract}

New PKS1830-211 radio frequency observations of methanol at a redshift of 0.88582
have established the most stringent limits on changes in the proton to electron
mass ratio $\mu$ to date. The observations place the limit of $\delta \mu/\mu
\le (0.0 \pm 1.0) \times 10^{-7}$ which is approximately a factor of four lower
than the previous lowest limit at a redshift of 0.6742. This stringent
limit at a look back time of roughly half the age of the universe has profound
implications for rolling scalar field cosmologies and the new physics that they
require.  Many of these cosmologies invoke a scalar field $\phi$ that is also 
coupled to the electromagnetic field causing the values of the fundamental 
constants, $\mu$ and the fine structure constant $\alpha$ to roll with time.
If the lowest expected value of the coupling to $\mu$, $\zeta_{\mu}$, is invoked
the new limit requires a limit on the dark energy equation of state parameter
$w$ such that $(w+1) \le 0.001$ at a redshift of 0.88582. This eliminates
almost all of the expected parameter space for such cosmologies and new physics
	\textbf{that have a coupling to the electromagnetic field.  In these cases}
the limit requires that $w$ must be extremely close to $-1$ for the
last half of the age of the universe or that the coupling of the rolling scalar
field to $\mu$ and the electromagnetic field be significantly below or at the 
limit of its expected range. The new observations solidify the role of fundamental 
constants in providing probes of the possible cosmologies and new physics to 
explain the acceleration of the expansion of the universe.

\end{abstract}


\section{Introduction} \label{s-intro}
Tracking the values of the fundamental constants such as the proton to electron
mass ration $\mu$ and the fine structure constant $\alpha$ has been shown to be
an effective way of discriminating between different mechanisms for the acceleration
of the expansion of the universe (\citet{thm12}, \citet{thm13}, hereinafter T12 and
TMV13). In particular
rolling scalar field cosmologies where the rolling field also couples with the
electromagnetic field predict changes in the values of $\mu$ and $\alpha$. Comparison
of the predicted changes in $\mu$ with the observed constraints translates to
equivalent constraints on the cosmologies and the new physics required by the
cosmology.

A new constraint on the change in $\mu$ at a redshift of 0.89 has just been
published by \citet{bag12}. They find $\Delta \mu / \mu = (0.0 \pm 1.0)
\times 10^{-7}$ in the radio absorption spectrum of PKS 1830-211 at a redshift 
of 0.88582.  This more than a factor of six improvement over the previous limit
of $\Delta \mu / \mu = (0.0 \pm 6.3) \times 10^{-7}$ in the same object 
\citep{ell12} and almost a factor of 4 improvement over the previous lower limit 
at a redshift of 0.6847 \citep{kan11}.  The $3\sigma$ constraint is based on 4 
methanol absorption 
lines observed with the 100m Effelsberg radio telescope.  Two of the absorption
lines are normal rotational transitions while the two other transitions are
mixed torsion-rotation transitions.  The sensitivity constant $K_{\mu}$, defined by
$\Delta \nu / \nu = K_{\mu} \times \Delta \mu / \mu$ for the two pure rotational
transitions is -1 while the two torsion-rotation transitions have sensitivity
constants of - 7.4 and -32.8. The differences in sensitivity constants provides
a means of testing for a change in $\mu$ using transitions in a single molecular
species rather than comparison between two or more molecular species.

\section{Constraints on Rolling Scalar Field Cosmologies}
\textbf{The connection between rolling scalar fields and changes in the values
of the fundamental constants has been considered for more than 10 years,
eg. \citet{chi02} and \citet{dva02}.}
It is assumed in most rolling scalar field cosmologies that the rolling
scalar field $\phi$ is also coupled with the electromagnetic field \citep{cop04}.
This coupling then alters the values of the fundamental constants as the
field changes with time (rolls).  This has been discussed by \citet{nun04}, 
hereinafter NL4
in the context of a linear coupling
\begin{equation} \label{eq-dmu}
\frac{\Delta x}{x} = \zeta_x \kappa (\phi - \phi_0)
\end{equation}
where $x = \mu, \alpha$, $\zeta$ is the coupling constant and 
$\kappa = \frac{\sqrt{8 \pi}}{m_p}$ where $m_p$ is the Planck mass. 
The same physics applies to both $\mu$ and $\alpha$ and the two are related by
\begin{equation}  \label{eq-amu}
\frac{\dot{\mu}}{\mu} \sim \frac{\dot{\Lambda}_{QCD}}{\Lambda_{QCD}} -\frac{\dot{\nu}}
{\nu} \sim R \frac{\dot{\alpha}}{\alpha}
\end{equation}
where $\Lambda_{QCD}$ is the QCD scale, $\nu$ is  the Higgs vacuum expectation 
value and R is a scalar often considered to be on the order of -40 to -50 
\citep{ave06}. NL4 further show the relationship between the rolling
field and the equation of state parameter $w \equiv \frac{p_{\phi}}{\rho_{\phi}} = 
\frac{\dot{\phi}^2 - 2 V(\phi)}{\dot{\phi}^2 + 2 V(\phi)}$ as
\begin{equation} \label{eq-wa}
w + 1 = \frac{(\kappa \phi')^2}{3 \Omega_{\phi}}
\end{equation}
where $\Omega_{\phi}$ is the dark energy density. Here $\dot{\phi}$ and $\phi'$ indicate 
differentiation with respect to cosmic time and to $N = \log{a}$ respectively where $a$ is 
the scale factor of the universe.  T12 and TMV13 then showed the relation
between changing fundamental constants and the evolution of $w$.
\begin{equation} \label{eq-wmu}
(w +1) = \frac{(\mu'/\mu)^2}{3 \zeta_{\mu}^2 \Omega_{\phi}} = \frac{(\alpha'/\alpha)^2}{3 \zeta_{\alpha}^2 \Omega_{\phi}}
\end{equation}
For the case of the proton to electron mass ratio Equation~\ref{eq-wmu} establishes the connection between
the product of a new physics parameter $\zeta_{\mu}^2$ and a cosmological parameter $(w+1)$ such that
\begin{equation} \label{eq-lim}
(w+1)\zeta_{\mu}^2 = \frac{(\mu'/\mu)^2}{3 \Omega_{\phi}} =\frac{(\Delta \mu/\mu)^2}{3 z^2 \Omega_{\phi}}
\end{equation}
where z is the redshift. \textbf{In creating Figures~\ref{fig-thmplot} and~\ref{fig-small}
equation~\ref{eq-omega} for the dark energy density is used. In section~\ref{s-examp}
dark energy densities are calculated using equations specific to the particular
cosmology.
\begin{equation}\label{eq-omega}
\Omega_{\phi}(a) = [1+(\Omega_{\phi 0}^{-1} - 1)a^{-3}]^{-1}
\end{equation}
}

The observational limits on $\Delta \mu/\mu$ therefore impose limits on the $(\Delta \mu/\mu)^2$ - $(w+1)$
parameter space that a rolling scalar field cosmology can live in.  This parameter space was first discussed
in T12, however, the new methanol observation has greatly reduced the available parameter space.

\section{The New Methanol Constrained Parameter Space} \label{s-met}

\begin{figure}
  \vspace{150pt}
\resizebox{\textwidth}{!}{\includegraphics[0in,0in][14.in,3.in]{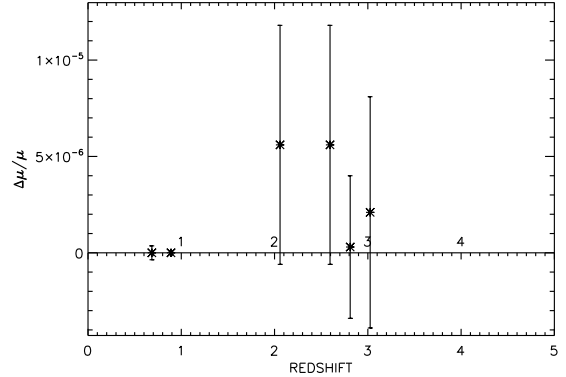}}
  \caption{The observed values of $\Delta \mu / \mu$ and their associated errors.
Note that the two lowest redshift (radio) errors are $3\sigma$ errors while the 
rest (optical) are $1\sigma$ error bars.} \label{fig-con}
\end{figure}

Figure~\ref{fig-con} shows the existing constraints on $\Delta \mu/\mu$ where the most stringent 
constraint for each object is indicated.  The radio results are $3\sigma$ constraints while the optical
constraints are $1\sigma$ measurements.  All of the measurements are from the references in Table 1
of T12 except for the new PKS 1830-211 measurement by \citet{bag12}.
It is obvious from Fig.~\ref{fig-con} that the current radio results are the most restrictive
in terms of $\Delta \mu/\mu$ and provide a condition at $z = 0.88582$ that must be satisfied
by any cosmology.

\begin{figure}
  \vspace*{75pt}
\resizebox{\textwidth}{!}{\includegraphics[0in,0in][14in,3.in]{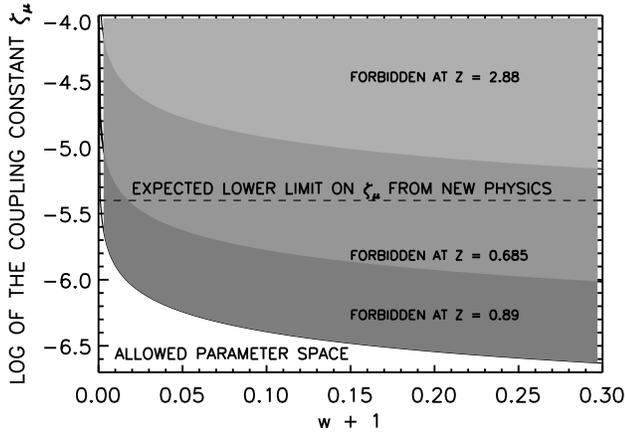}}
  \caption{The figure shows the forbidden and allowed parameter space in the $\zeta_{\mu}$,
$(w+1)$ plane based on the three most restrictive low and high redshift
observations. The upper light shaded area is for the constraint at a redshift of 2.811,
the middle darker area and above are for a redshift of 0.685, and lower dark shaded 
area and above is for the new constraint at a redshift of 0.89.   
The dashed line indicates the lower expected limit on the coupling 
factor $\zeta_{\mu}$ as discussed in the text.} \label{fig-thmplot}
\end{figure}

Figure~\ref{fig-thmplot} shows the effect of the new measurement on the $(\Delta \mu/\mu)^2$ - $(w+1)$
parameter space along with the other radio restriction at $z = 0.6847$ \citep{kan11} and the most
restrictive optical measurement at $z = 2.811$ \citep{kin11}.
The parameter space diagram in Fig.~\ref{fig-thmplot} is similar to the equivalent
diagram in T12 with different constraints.  The shaded areas indicates the parts 
of the diagram
that are forbidden.  All of the area above the bottom line are forbidden at the redshift listed
just above the line.  All of the light shaded area and above is forbidden by the constraint on
$\Delta \mu/\mu$ at $z = 2.88$ and all of the shaded areas and above are forbidden 
by the new $\Delta \mu/\mu$ measurement at $z = 0.89$.  

The dashed line in Fig.~\ref{fig-thmplot} indicates the expected lower limit 
on $\zeta_{\mu}$ from new physics. 
\textbf{\citet{nun04} use  the work of \citet{cop04} to set limits on the expected value
of the coupling with the fine structure constant $\zeta_{\alpha}$ of
$10^{-7} \le \zeta_{\alpha} \le 10^{-4}$.  With the expected value of $-40$ for R 
in Eq.~\ref{eq-amu} this produces the expected range for $|\zeta_{\mu}|$ of
$4 \times 10^{-6} \le \zeta_{\mu} \le 4 \times 10^{-3}$.  The lower bound from
\citet{nun04} was determined by setting $\Delta \alpha/\alpha = 10^{-7}$, about
a factor of 100 below the reported value of $\sim 10^{-5}$ by \citet{kin12}.}

 It is
always possible to accommodate any value of $(w+1)$ by lowering the value of $\zeta_{\mu}$, and of
course the standard model of physics expects $\zeta_{\mu}$ to be zero.  This, however, is a test
of non-standard models with rolling scalar fields that do interact with the electromagnetic field.
It is obvious from Fig.~\ref{fig-thmplot} that if we set $\zeta_{\mu}$ at it lowest expected value
of $4 \times 10^{-6}$ then, at a redshift of 0.89, the value of $w$ is very close to $-1$.  Since 
the value of $\Delta \mu$ is the difference between the value of $\mu$ today and its value at
$z=0.89$ this implies that $w$ has been very close to $-1$ for the entire time between now and
$z=0.89$ unless the scalar field deviates from its value at $z=0.89$ and then returns to that
value at the present time.  Most rolling scalar field models do not exhibit this behavior so
they must conform to having $w$ close to $-1$ for roughly half the age of the universe.

\begin{figure}
  \vspace*{75pt}
\resizebox{\textwidth}{!}{\includegraphics[0in,0in][14in,3.in]{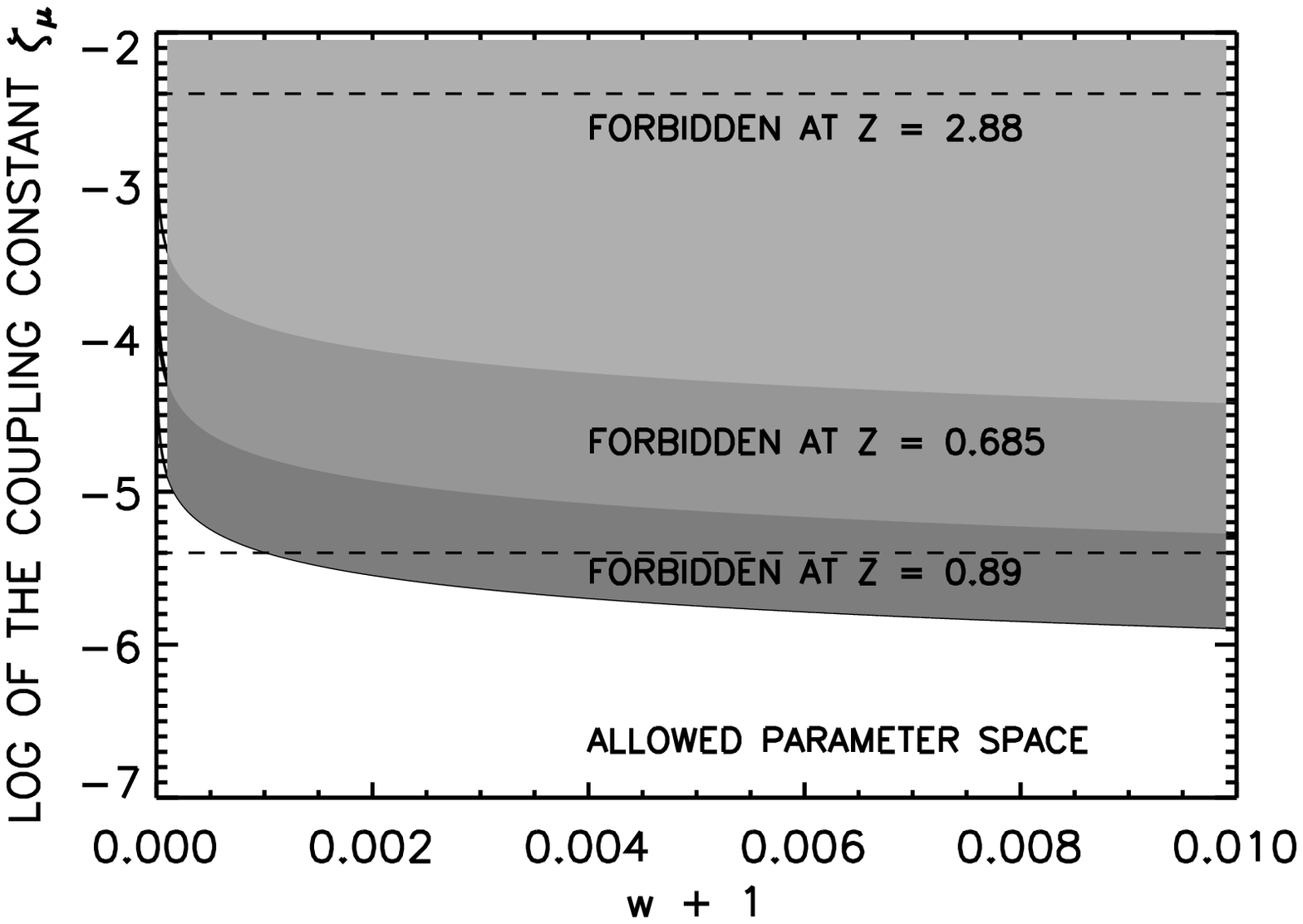}}
  \caption{The figure gives a detailed view of the narrow allowed limits on $w+1$ if the
value of $\zeta_{\mu}$ is taken at its lower expected limit shown by the lower dashed
line. At a redshift of 0.89 $w+1$ is constrained to be less than $0.001$ unless the
coupling constant is reduced below its lowest expected value.} \label{fig-small}
\end{figure}

To better quantify the restriction Fig~\ref{fig-small} shows an expanded view of the region 
near $(w+1) = 0$.
Figure~\ref{fig-small} indicates that if we hold to the lowest expected value of $\zeta_{\mu}$
we must place the limit of $(w+1) \le 0.001$ for all redshifts of $0.89$ or less. This is an
extreme restriction that puts the validity of current models of rolling scalar fields
\textbf{that also couple with the electromagnetic field} as the 
driving force for the acceleration of the expansion of the universe in doubt. The results are,
however, consistent with a $\Lambda$CDM universe and the standard model of physics. Some
areas for possible modification of the current rolling scalar field models are
discussed in Section~\ref{s-ar}.

\section{Examples for Individual Sample Rolling Scalar Fields} \label{s-examp}
\begin{figure}
  \vspace*{75pt}
\resizebox{\textwidth}{!}{\includegraphics[0in,0in][14in,3.in]{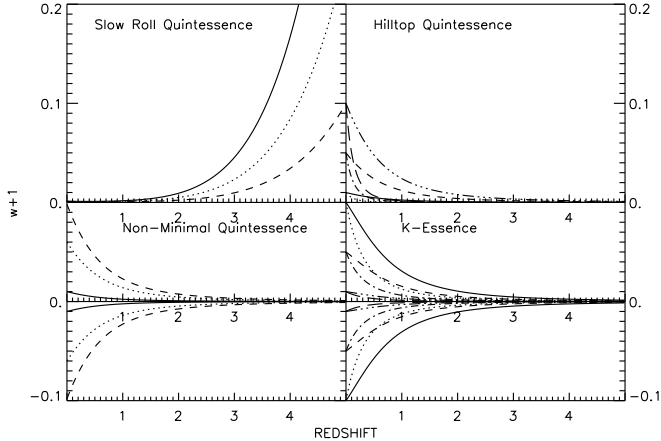}}
  \caption{The figure shows the evolution of the equation of state parameter $w$
by plotting the value of $w+1$ as a function of redshift for each of the four
cosmologies. The last column of Table 2 in T12 contains the line
style code for each of the cases.} \label{fig-w}
\end{figure}

\begin{figure}
\vspace{100pt}
\resizebox{\textwidth}{!}{\includegraphics[0in,0in][14in,3.in]{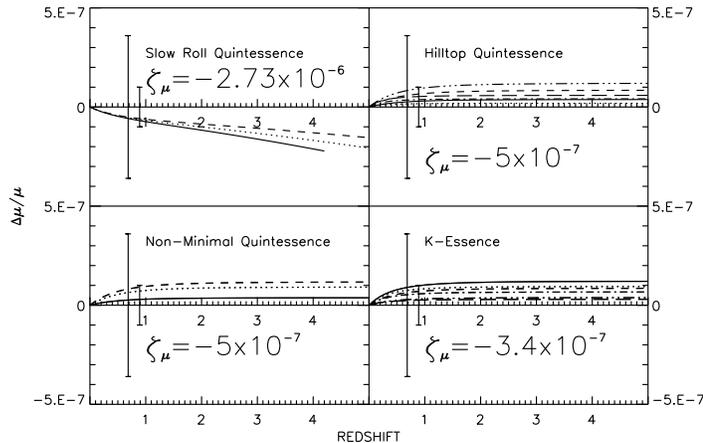}}
  \caption{This figure plots the evolution of $\Delta \mu / \mu$ versus redshift for
each of the four cosmologies. The value of $\zeta_{\mu}$ has been adjusted in each
cosmology so that all cases for that cosmology fall within the observational 
constraints. The higher redshift constraint at z=2.811 is not plotted since it
is larger than the plot size. The required value of $\zeta_{\mu}$ to meet the
constraint is printed in each plot. Refer to Table 2 of T12 
for the line style for each case.} \label{fig-fit}
\end{figure}

The $\Delta \mu/\mu$ constraints at the various redshifts provide parameter space "wickets"
through which valid cosmologies must pass.  Since all of the current "wickets" include the
space where $\Delta \mu/\mu = 0$, $\Lambda$CDM with the standard model of physics clears all
of them as do most modified gravity theories.  Rolling scalar field models with non-standard
physics do not necessarily pass the test.  TMV13 investigated some quintessence and
K-Essence models to determine the parameter space that fit the observations, where in this
case the parameter space included those individual parameters that define the particular
model such as an initial value of $w$ at a given value of redshift.  In the following the
models examined in TMV13 are used with the new restriction and the same 
individual model parameters with only $\zeta_{\mu}$ being varied to achieve a 
fit.  As described in TMV13 the value of $\Delta \mu/\mu$ at any scale 
factor $a$ is given by
\begin{equation} \label{eq-muint}
\frac{\Delta\mu}{\mu} =\zeta_{\mu}\int^{a}_{1}\sqrt{3\Omega_{\phi}(x)(w(x)+1)}x^{-1}dx
\end{equation}
with $\Omega_{\phi}(x)$ and $(w(x)+1)$ determined by the details of the particular 
cosmology. Table 2 of TMV13 lists the values of the parameters used in each
of the cosmologies.  Note that the matches will be slightly different in the individual
cases than in the general case discussed above since in the general case we use 
Equation~\ref{eq-omega} for $\Omega_{\phi}(x)$ whereas in the individual 
cases we use the more accurate
\begin{equation}\label{eq-omegaexp}
\Omega_{\phi}(a) = [1+(\Omega_{\phi 0}^{-1} - 1)a^{-3} exp(3\int^{a}_{1}
\frac{(1+w(x)}{3}dx]^{-1}
\end{equation}
Figure 5 from TMV13 plots the values of $w + 1$ versus redshift for the 4
cosmologies, slow roll quintessence, hilltop quintessence, non-minimal quintessence,
and k-essence, and is repeated here as Fig.~\ref{fig-w} for ease of reference.
It is obvious from Fig.~\ref{fig-w} that none of the three thawing cosmologies come
close to satisfying the limit of $(w+1) \le 0.001$ at a redshift of $0.89$.
Although not obvious from the scale of the plot even the freezing slow roll
quintessence cosmology does not satisfy the constraint. Since the constraints are
in a parameter space that has both a cosmology and a new physics component, solutions
can be sought in either area.  The easiest is to simply adjust the new physics
parameter $\zeta_{\mu}$ downward, lower than the expected value, to open up the
$w+1$ parameter space.  The results of doing this are presented in Fig.~\ref{fig-fit}.
The required value of $\zeta_{\mu}$ needed to make all of the cases fit the 
constraints is shown in each figure.  In some cases, such as K-Essence the
required value of $\zeta_{\mu}$ is more than a factor of $10$ below the expected
lower limit.

\section{Areas for the Recovery of Viable Rolling Scalar Field Models} \label{s-ar}
Quantitative and definitive new models for rolling scalar fields are beyond the 
scope of this paper and will be reserved for future papers.  There are, however,
obvious areas for modification which are discussed briefly below.

\subsection{Varying $\zeta_{\mu}$ and or R} \label{ss-vr}
\textbf{The lower bound on $\zeta_{\mu}$ as discussed in Section~\ref{s-met} is relatively
arbitrary, therefore, the most straight forward method to recover viable rolling
scalar field models is to simply reduce the value of $\zeta_{\mu}$ by a factor
on the order of 10.  This recovers most of the $(w+1)$ parameter space appropriate
for reasonable cosmological models. This can be accomplished by either making
the two terms in equation~\ref{eq-amu} roughly equal to each other or by
lowering the time variation of each of the terms.  The former is equivalent to
reducing the value of R while the latter reduces all of the coupling between
the rolling scalar field and the electromagnetic field.  Further discussion of
this point is beyond the scope of the present paper.}

\subsection{The Form of the Coupling to the Electromagnetic Field}
\citet{nun04} use a simple linear form of coupling between the electromagnetic
field and the scalar field (Equation~\ref{eq-dmu}). This is quite reasonable given
the lack of information on the true form of the coupling and can be considered
as the first term in a Taylor series expansion of the actual coupling.  The 
coupling, however, may be of a quite different form which could lower the effect
on the electromagnetic field from the expected range. The coupling has also been
considered to be constant with time which may also be an erroneous assumption.

\section{Summary}
The new limit on $\Delta \mu/\mu$ of $(0.0 \pm 1.0) \times 10^{-7}$ by \citet{bag12}
places significant constraints on rolling scalar fields as the origin of the
acceleration of the expansion of the universe.  The measured invariance of $\mu$
should be considered in the formulation of cosmologies invoking rolling scalar fields.
Current results are consistent with acceleration due to a cosmological constant
$\Lambda$ and the standard model of physics. If the expected limits on the 
coupling constant $\zeta_{\mu}$ are accepted then the invariance at a redshift
of 0.89 implies that the value of the equation of state variable $w$ has been
within $0.001$ of $-1$ for the last half of the age of the universe. \textbf{If, however,
the coupling between the rolling scalar field and the electromagnetic field is
weaker than expected then the constraints on $w$ are significantly relaxed.  This,
on the other hand, has implications for new physics.}  More accurate 
measurements of $\mu$ at higher redshift will be very useful in the further 
consideration of rolling scalar fields as the origin of the late time acceleration
of the expansion of the universe.

\section{Acknowledgments}
The author wishes to thank Christian Veillet for bringing the paper \citet{bag12}
to his attention. The author also thanks Carlos Martins and Paolo Molaro for 
illuminating discussions on the general topic of varying fundamental constants.
\textbf{The paper was also improved by helpful comments from an anonymous referee.}

\label{lastpage}
\end{document}